\documentclass[12pt]{article}
\renewcommand{\baselinestretch}{1.2}
\usepackage{amsmath}
\usepackage{amssymb}

\usepackage{indentfirst}
\usepackage{amssymb}
\usepackage{amsfonts}
\usepackage{amscd}
\usepackage{amsbsy}
\usepackage{amsthm}
\usepackage{latexsym}
\usepackage{graphicx,color} 
\usepackage[dvipsnames]{xcolor}
\usepackage[colorlinks]{hyperref}
\hypersetup{linkcolor=blue,citecolor=blue,urlcolor=blue}

\usepackage{mathrsfs}
\usepackage{mathtools}
\usepackage{physics}
\usepackage{comment}
\usepackage[normalem]{ulem} 

\def\nn{\nonumber}       
\def\beq{\begin{eqnarray}}
\def\eeq{\end{eqnarray}}

\def\ln{\,\mbox{ln}\,}

\def\diag{\,\mbox{diag}\,}

\def\al{\alpha}
\def\be{\beta}

\def\ga{\gamma}
\def\de{\delta}
\def\vp{\varphi}

\def\ka{\kappa}
\def\la{\lambda}
\def\na{\nabla}
\def\pa{\partial}

\def\si{\sigma}
\def\om{\omega}
\def\ph{\varphi}

\def\Ga{\Gamma}
\def\De{\Delta}
\def\La{\Lambda}

\DeclareMathOperator{\cx}{\square}

\newcommand{\n}{\nabla}

\newcommand{\tx}{\text}

\newcommand{\cl}{\mathcal}

\newcommand{\bpsi}{\bar{\psi}}

\usepackage{float} 
\usepackage{titlesec}

\titleformat*{\section}{\large\bfseries}
\titleformat*{\subsection}{\normalsize\bfseries}

\date{\today}

\begin{document}

\begin{center}

{\Large 
Trace anomaly, effective approach, and gravitational potential}
\vskip 6mm

{\bf Riccardo Fecchio}$^{a,b}$
\hspace{-1mm}\footnote{E-mail address: \ riccardo.fecchio@edu.ufes.br},
\ {\bf Ilya L. Shapiro}$^{b,a}$
\hspace{-1mm}\footnote{E-mail address: \ ilyashapiro2003@ufjf.br},
\vskip 6mm

$^{a}$ 	PPGCosmo,
Universidade Federal do Esp\'{\i}rito Santo
\\
Vit\'oria, 29075-910, ES, Brazil
\vskip 2mm

$^{b}$ Departamento de F\'{\i}sica, ICE,
Universidade Federal de Juiz de Fora
\\
Juiz de Fora, 36036-900, Minas Gerais, Brazil

\end{center}
\vskip 2mm
\vskip 2mm

\centerline{\large{\bf Abstract}}
\begin{quotation}

\noindent
We explore and discuss corrections to the Newton potential
from the quantum effects of conformal matter fields. In this special
case, one can compare different approaches, including that of
effective quantum gravity and another, based on the conformal (trace)
anomaly. The comparison of these two methods is the main focus in
the present work. Using the anomaly-induced effective action of
gravity requires fixing the quantum vacuum state, similar to what
is done in the description of black hole evaporation. In the Boulware
vacuum state, we compute the anomaly-induced stress tensor and the
first-order correction to the classical gravitational law. The quantum
correction to the Newton's potential derived in this way differs
from the result calculated in a way analogous to the effective
approach to quantum gravity. The only way to reconcile the two
approaches for deriving the leading semiclassical corrections to the
Newtonian potential is to modify the asymptotic behavior of the
average of the energy-momentum tensor in the Boulware vacuum
state, as has been recently discussed in the literature.
\vskip 3mm

\noindent
\textit{Keywords:} Quantum corrections, Schwarzschild solution,
effective approach, conformal anomaly, Boulware vacuum
\vskip 3mm

\noindent
\textit{MSC:} \
81T50,  
81T20,   
83C47,   
83C57  

\end{quotation}


\newpage
\tableofcontents

\section{Introduction}
\label{sec1}

One of the main motivations for quantum and semiclassical gravity
theories comes from the high-energy physics. As examples, one can
mention the Starobinsky model of inflation \cite{star} or creation of
particles from vacuum in the reheating after inflation (see, e.g.,
\cite{ParToms}). Another example of creating particles from the
vacuum is the Hawking radiation \cite{Hawking75}. It is well-known
that this effect can be attributed to the conformal (trace) anomaly
\cite{chfu}. Moreover, one can obtain more detailed information using
the anomaly-induced, {vacuum effective action \cite{rie,frts84}. The
local representation of this action involves two auxiliary scalar
fields \cite{a}. Changing the boundary conditions for these fields,
one can reproduce the classification of the vacuum states around a
black hole \cite{balsan}, which was originally obtained by different
approaches. In particular, it is possible to reproduce the asymptotic
behaviour of the mean value of the energy-momentum tensor in the
Boulware vacuum state, which is appropriate for describing vacuum
quantum effects for an observer situated far from the horizon, in
the asymptotically Minkowski spacetime. In the classical theory,
this corresponds to the weak-gravity regime, with the deviation from the
flat metric proportional to the Newtonian gravitational potential.
In other words, the trace anomaly offers a way for deriving the
gravitational potential, by mimicking the Boulware vacuum and
approximating the mean value of the energy-momentum tensor.


In general, one can expect that the remnants of semiclassical or
quantum gravity may persist in the weak gravitational fields. This
expectation represents a
motivation for the efforts to measure possible deviations from
Newton's law of gravity. The precise experiments in this area are the
subject of great efforts and interest (see, e.g., \cite{Homans2025}).

From the theoretical side, in general relativity (GR) and field theory,
there are several ways to derive Newton's gravitational law. A
standard approach is to take the long-distance limit of the
Schwarzschild solution; another one is to use the tree-level Feynman
diagram with a one-graviton exchange between two masses. In GR,
the results are the same. However, the situation becomes more
complicated in modified gravity theories, especially in theories
with massive ghosts. One example where this issue has been well
studied is light bending and corrections to the Newtonian potential
in six-derivative gravity, where the use of Feynman diagrams may
produce misleading results \cite{ABS}.

A different situation arises in the standard effective approach (see
\cite{Handbook-eff} for the set of reviews),
where higher-derivative terms are regarded, by definition, as small
corrections to the Einstein-Hilbert action \cite{Simon-90}. In
quantum gravity, the effective approach produces
$\mathcal{O}(1/r^3)$-type corrections to the Newtonian potential
\cite{don}. The coefficient depends on the calculation scheme, especially
on whether the matter source of gravity is quantized or not. In the first
case, one can deal with $S$-matrix elements \cite{BBDH} (see also
\cite{KirKhri}), which
guarantee gauge-fixing independence when the complete set of diagrams
is included \cite{SQED}.   If the matter (macroscopic massive bodies)
is not quantized, the corresponding diagrams should be omitted. Then
the situation is more complicated since the framework is a quantum
field theory with external conditions and the $S$-matrix approach is
not directly applicable. In this case, an elegant scheme to ensure
gauge-fixing independence has been proposed in \cite{DM97}. This
scheme, which we shall partially use below, consists of deriving
quantum corrections to the effective action and, after that, of
considering the motion of a test particle. The gauge-fixing independence
follows from the fact that the result is an on-shell combination
of beta functions \cite{EffQG-TNLM}.

The purpose of the present work is to check the
correspondence of two approaches, namely the one based on the
anomaly-induced action \cite{balsan}, and the one based on the
effective field theory methods \cite{don}. The idea is to derive the
semiclassical corrections to the Newtonian potential in two ways
and compare the results. Similar considerations were previously
given in Refs.~\cite{ABF} and \cite{BoassoMaz2025}, but we
arrive at similar results in a different way and make the
existing contradiction more explicit.

Let us anticipate that the effective field theory scheme
will be applied to massless conformal matter fields only. We
stress that it is important to considerer the conformal fields because
we intend to make a comparison with the method based on the
trace anomaly. In the trace anomaly approach, the modified
Newtonian potential
comes from the quantum-corrected Schwarzschild solution.
Such a solution has been the object of investigations for a long time, see
e.g., \cite{Duff74Sch}. Here we apply the approach that combines
the anomaly-induced action \cite{rie,frts84}, which enables one
to deal with the leading logarithms of the one-loop contributions
and the effective approach, treating quantum terms as small
perturbations, according to \cite{Simon-90}.
The directly calculated one-loop contributions of
conformal fields include form factors such as $\log(\cx/\mu^2)$.
The anomaly-induced action enables one to replace these form factors
with simpler Green functions of the fourth-derivative
Paneitz operator. Furthermore, one can reformulate the induced
action in a local form using one \cite{rie} or two \cite{a} auxiliary
scalar fields. In the latter case, the effective action preserves all the
information about the UV part of loop corrections and, for this
reason, this form of the effective action is most useful for the
classification of vacuum states of conformal fields on the background
of the classical Schwarzschild solution \cite{balsan}.

The quantum corrections to a given classical solution may depend
on the choice of the vacuum state of quantum fields, and the latter
is well-known to be ambiguous in curved spacetime \cite{Fulling73}.
On the other hand,
the form of logarithmic corrections is fixed by the UV divergences
and does not depend on the choice of vacuum state for the quantum
fields. The same is true for the form of the
anomaly-induced action,
which is a handy form to parameterize logarithmic corrections.
On the other hand, the classification of vacuum states is performed by
fixing boundary conditions for the auxiliary fields, which is equivalent
to specifying the boundary conditions for the Green functions in the
nonlocal form of the action \cite{balsan}. As we shall see below, the
analysis of quantum corrections to the Schwarzschild solution also
requires choosing these boundary conditions or, equivalently, the
vacuum state of the quantum fields. In this work, we use the Boulware
vacuum for deriving the solution far from the black hole, i.e., for
$r \gg r_g$ with $r_g$ the Schwarzschild radius. We will see that
the correction to the Newton gravitational law, obtained by this
approach, is surprisingly different from the one calculated using
diagrams, as was already noted previously in \cite{ABF}.

The paper is organized as follows. In Sec.~\ref{sec2} the scheme for
deriving one-loop effective corrections to the Newtonian potential in
quantum gravity, used in \cite{DM97} and \cite{EffQG-TNLM}, is
adapted for the contributions of quantum conformal matter. In
Sec.~\ref{sec3} the anomaly-induced effective action and the
corresponding stress-tensor are introduced. In Sec.~\ref{sec4}
corrections to the Newtonian potential are derived from the trace
anomaly, using the objects introduced in Sec.~\ref{sec3}.
Finally, in Sec.~\ref{sec5} we draw our conclusions and present
final discussions.

\section{Semiclassical Newton potential in the effective approach}
\label{sec2}

Our starting point in this section will be the free massless
conformal theory of $N_s$ scalar, $N_f$ fermion and $N_v$ vector
fields in an arbitrary curved background,
\begin{align}
\label{eq:conformal-action}
\cl{S} \,=\,
\int \dd[4]{x}\sqrt{-g}\,\,
\bigg\{ \sum^{N_s}_{i=1} \frac12 \Big[ (\pa\phi_i)^2
+ \frac{1}{6} R \phi_i^2\Big]
+ \sum^{N_f}_{j=1}\bpsi_j \big(i \ga^\mu \na_\mu \big) \psi_j
+ \sum^{N_v}_{k=1} {F^k_{\mu\nu}}^2 \bigg\} \,,
\quad
\end{align}
where $F^k_{\mu\nu} = \n_\mu A^k_\nu -\n_\nu A^k_\mu$. The
$N_v$ copies of an Abelian vector field can be traded for the same
number of Yang-Mills fields (not multiplets) without changing the
result for the vacuum contributions. The coefficient of the
nonminimal coupling $R\phi^2$ is taken to be $1/6$, making
the action invariant under local conformal transformations.

The one-loop divergences in the vacuum effective action are
pretty well known \cite{birdav} (see also \cite{OUP} for a
textbook-level introduction and details), and have the form
\begin{align}
\bar{\Gamma}^{(1)}_{div}
\,=\,
-\,\frac{\mu^{n-4}}{n-4}
\int \dd[n]{x}\sqrt{-g} \,\,
\Big( \be_1 C^2 + \be_2 E_4 + \be_3 \cx R \Big) \,,
\qquad
\be_l \,=\,\frac{d_l}{(4\pi)^2}\,,
\label{Gadiv}
\end{align}
where $C^2$ and $E_4$ are the square of the Weyl tensor and the
integrand of the Gauss-Bonnet term, and $n$ is the parameter of
dimensional regularization. The coefficients are \cite{birdav,OUP}
\begin{subequations}
\label{k123}
\begin{align}
& d_1 = \frac{N_s}{120} +\frac{N_f}{20} +\frac{N_v}{10}\,,
\\
& d_2 = - \frac{N_s}{360} - \frac{11}{360}N_f - \frac{31}{180} N_v\,,
\\
& d_3 = \frac{N_s}{180} +\frac{N_f}{30} -\frac{N_v}{10}\,.
\end{align}
\end{subequations}
We note that the absence of the term $R^2$ is due to
conformal symmetry. It is also important for us that the coefficients
$d_{1,2,3}$ do not depend on the gauge fixing for the gauge field.
These two points are part of the difference from the case  of
effective quantum gravity with the classical source given by point
particles with the energy-momentum $T_{\mu\nu}$. In this case,
the expression for the divergences is
\beq
\bar{\Ga}^{(1)}_{\text{div}}
&=&
- \,\,\frac{\mu^{n-4}}{n-4} \int \dd[n]{x} \sqrt{-g} \, \Big\{
\beta_W C^2 - \frac13 \beta_{R2} R^2
-  2 \ka^2 \beta_{\scriptscriptstyle RT1} R_{\mu\nu} T^{\mu\nu}
\nn
\\
&&
\quad
+ \,\, \ka^2\beta_{\scriptscriptstyle RT2} R T
+ \ka^4 \beta_{\scriptscriptstyle TT} T^2 \Big\},
\label{divWeyl}
\eeq
In the case of effective quantum gravity, all individual beta functions
in (\ref{divWeyl}) depend on the gauge fixing and also on the
parametrization of the quantum metric. Only the unique special
combination (derived using the on-shell condition)
\beq
\label{betainv}
\beta_{\text{inv}}
\,=\,
\frac43 \beta_W - \frac13 \beta_{R2} - 2\beta_{\scriptscriptstyle RT1}
- 2\beta_{\scriptscriptstyle RT2} + 4 \beta_{\scriptscriptstyle TT}
\eeq
is a gauge and parametrization invariant quantity \cite{EffQG-TNLM}.
One can easily rewrite (\ref{Gadiv}) by noting that the terms $E_4$
and $\cx R$ are irrelevant for the one-graviton exchange between
two massive particles.
Thus, (\ref{Gadiv}) can be regarded as a simpler version of
(\ref{divWeyl}), with a unique nonzero beta function $\be_W = \be_1$
and hence $\be_{\text{inv}} = \frac43 \be_W$
which does not depend on the gauge fixing. 

One of the differences between effective quantum gravity and the
semiclassical case is that 
the latter can be made renormalizable by
introducing the complete vacuum action with the terms quadratic
in curvatures,\footnote{Effective approach to quantum gravity
ignores its non-renormalizability.}
\begin{align}
\label{classical-action}
\cl{S}_{\tx{grav}}
\,=\, - \frac{1}{\kappa^2} \int\dd[4]{x}\sqrt{-g} R
\,+ \int\dd[4]{x}\sqrt{-g}\,\,
\big( a_1C^2 +a_2E_4 +a_3\square R \big)\,,
\end{align}
where $\ka^2=16\pi G$ and $a_i$ are higher-derivative coefficients.
As in the divergences (\ref{divWeyl}), only the constant
$a_1 = - \la/2$ is relevant for our purposes.

Summing up the relevant terms in (\ref{Gadiv}) and
(\ref{classical-action}), we trade the divergence for the leading
logarithm according to the rule, 
\beq
\frac{\mu^{n-4}}{n-4}
\,\,\,\,\longrightarrow \,\,\,\,
\frac12 \ln \Big(\frac{\cx}{\,\,\mu^2}\Big)\,.
\label{div2log}
\eeq
This identification is a pretty standard thing (see, e.g., 
\cite{DM97} and references therein); the detailed derivation
can also be found in \cite{OUP}. In brief, this formula reflects the 
proportionality relation between the leading logarithm in the 
nonlocal formfactor and the logarithmic divergence, which is 
mapped into $1/(n-4)$ in dimensional regularization. Another 
important aspect is the relation between the divergence and the anomaly
\cite{duff77}. Altogether, we arrive at the links between the leading 
logarithms, the $1/(n-4)$-pole in the divergences and the trace anomaly. 

Using (\ref{div2log}) we obtain the semiclassical analog of the 
effective quantum gravity expression for the finite part of the 
effective action \cite{DM97,EffQG-TNLM},
\beq
\label{Weyl-log}
\Ga^{(1)}_{\tx{fin}}
\,= \, -\, \frac12 \int \dd[4]{x}\sqrt{-g} \,\,
C_{\rho\si\alpha\beta} \bigg[  \la \,+\,
\be_1 \ln \Big(\frac{\cx}{\,\,\mu^2}\Big)
\bigg]C^{\rho\si\al\be}\,.
\eeq
The classical term in this expression can be absorbed into a
redefinition of $\mu$, and we get an expression which is a particular
form of the one considered in the effective quantum gravity case.
Thus, without loss of generality, we can set $\la=0$, such that
the expression (\ref{Weyl-log}) becomes a simplified version of
the corresponding expression from \cite{EffQG-TNLM},
\beq
&&
\bar{\Ga}^{(1)}_{\tx{fin}} \,=\,
- \int \dd[4]{x} \sqrt{-g} \,  \bigg\{
\frac12 \be_W C_{\mu\nu\al\be}
\ln \Big(\frac{\cx}{\,\,\mu^2} \Big) C^{\mu\nu\al\be}
- \frac{1}{6} \be_{R2} R \ln \Big(\frac{\cx}{\,\,\mu^2}\Big) R
\nn
\\
&&
\quad
- \, \ka^2 \beta_{\scriptscriptstyle RT1}  R_{\mu\nu}
\ln \Big(\frac{\cx}{\,\,\mu^2}\Big) T^{\mu\nu}
+ \frac12  \ka^2 \beta_{\scriptscriptstyle RT2}
\, R \ln \Big(\frac{\cx}{\,\,\mu^2}\Big) T
+ \frac12  \ka^4 \beta_{\scriptscriptstyle TT} T
\ln \Big(\frac{\cx}{\,\,\mu^2}\Big) T \bigg\}.
\label{EA-Weyl}
\eeq
Eq.~(\ref{Weyl-log})
does not have terms involving the trace $T$ of the classical
energy-momentum tensor of point particles. The reason is that the
free conformal matter fields are not coupled to the classical matter
sources through the corresponding stress-tensor. Thus, these fields 
contribute only to the quantum corrections in the gravitational 
sector. Hence, the beta functions related to the trace of the 
energy-momentum tensor vanish.
Also, in the case (\ref{Weyl-log}) of quantum conformal matter
fields, there is no logarithmic term associated with the square of the
scalar curvature, owing to the local conformal symmetry.  
We note that it is fairly easy to extend
the calculation to a general $\xi$. In this case, there is a nonzero
expression (assuming the same $\xi$ for all scalars)
\beq
\be_{R2}\,=\,
\frac{N_s}{2\,(4\pi)^2}\,\Big(\xi-\frac16\Big)^2.
\label{beta4}
\eeq
Let us note that the comparison with the anomaly-induced method
is possible only for $\xi=1/6$ because otherwise this action leads to
contradictions \cite{rie}.

The expressions (\ref{Weyl-log}) and (\ref{EA-Weyl}) can be
equivalently derived directly using Feynman diagrams or by using
the rule (\ref{div2log}) in the divergences obtained by the
heat-kernel method.

The next question is how to use these
expressions to arrive at the quantum-corrected version of the Newtonian
potential. We shall follow the approach of \cite{DM97} that boils
down to the universal expression (\ref{betainv}), even though in our case the beta function $\be_W$ is invariant by itself. The
common point is that the expression (\ref{betainv})
is gauge-fixing independent in both cases.

Consider the procedure of deriving the corrected gravitational potential
in the theory with gravity described by the effective action
$\cl{S}_{\tx{grav}} + \bar{\Ga}^{(1)}_{\tx{fin}}$. The starting
point of the scheme of \cite{DM97} and \cite{EffQG-TNLM} is
choosing a static, point-like mass source with the action
\begin{align}
\label{point-likeM}
\cl{S}_M
\,=\, -\,M \int \dd{s}
\,=\, -M\, \int \sqrt{g_{\mu\nu}\dd{x}^\mu \dd{x}^\nu} \,,
\end{align}
and consider the gravitational field produced by this static mass.
We assume the metric
\beq
g_{\mu\nu}\,=\, \diag\Big\{
1 + 2\Phi(r),\,  - \,\delta_{ij} [1 - 2\Psi(r)]\Big\},
\label{Nmetric}
\eeq
where $\Phi$ and $\Psi$ are weak functions of the spatial radius $r$
only. The last condition is due to the isotropy of the problem under
consideration and to the fact that we are looking for a static solution.
Since the
corrections to the gravitational potential should be small, the
consideration is restricted to the linear order in  $\Phi$ and $\Psi$.

Making a perturbative expansion in quantum corrections, after several
intermediate steps, we arrive at the result for the two potential functions
\cite{DM97}
\beq
&&
\label{phi}
\Phi(r)\, = \,  - \,\frac{\ka^2 M}{16 \pi r}
\,-\, \Big(\frac43 \beta_W - \frac13 \beta_{R2}
- \beta_{\scriptscriptstyle RT1}
- \beta_{\scriptscriptstyle RT2} \Big) \frac{\ka^4 M}{8 \pi r^3} \, ,
\\
&&
\label{Psi}
\Psi(r) \,=\,  - \,\frac{\ka^2 M}{16 \pi r}
-  \Big(\frac23 \beta_W + \frac13 \beta_{R2} - \beta_{\scriptscriptstyle RT1}
+ \beta_{\scriptscriptstyle RT2} \Big) \frac{\ka^4 M}{8 \pi r^3}.
\eeq
In the quantum gravity case, these expressions do not have a direct
physical meaning because both are gauge-fixing dependent. Indeed, in
our semiclassical problem, all betas except $\be_W$ and $\be_{R2}$
vanish and these two are universal in the sense described above, i.e,
they do not depend on the gauge fixing. Thus, we can stop at the above
result. However, let us follow the full prescription in all theories.
Thus, we introduce the action of a test particle with a small mass $m$
instead of the large mass $M$ in the source (\ref{point-likeM}).
We disregard the effect of the small mass on the gravitational field, and
consider the motion of the test particle in the field created by $M$, in
the theory with the gravitational action
\begin{align}
\label{eq:IR-vacuum-action}
\Ga^{(1)}_{\tx{IR}} \,=\,
\cl{S}_{\tx{EH}} + \bar{\Ga}^{(1)}_{\tx{fin}} \,,
\end{align}
where $\cl{S}_{\tx{EH}}$ is the Einstein-Hilbert action. We avoid
literal repetition of the calculation made in \cite{DM97}, and also
presented in full detail in \cite{EffQG-TNLM}, and give only the
final result. The classical
limit for the quantum-corrected geodesic of a test particle
of a unit mass $m=1$ gives the
Newtonian potential with the weak quantum correction
\beq
V(r) = - \frac{\ka^2 M}{16 \pi r}
\,-\,
\frac{\ka^4 M}{8 \pi r^3 } \, \be_{\rm inv} .
\label{pot_final}
\eeq
In the semiclassical case, we have
\beq
\be_{\rm inv}
\,=\, \frac43\,\be_W \,-\, \frac13\,\be_{R2},
\label{betinv}
\eeq
where $\be_W = \be_1$ from Eqs.~(\ref{Gadiv}) and (\ref{k123}),
while the second term on the \textit{r.h.s.} vanishes in the
conformal limit and is defined by Eq.~(\ref{beta4}) otherwise.
The origin of the formula (\ref{pot_final}) is that the loops of
massless fields develop IR divergences and the finite distance
$r$ between the two point-like masses plays the role of a
natural regulator of these divergences in the long-distance limit.
As it has to be from this perspective, the formula is rather general,
e.g., it can be applied to effective quantum gravity \cite{DM97},
and, as we just saw, to conformal and non-conformal massless matter
fields. In all cases, quantum corrections are $\mathcal{O}(r^{-3})$.

It is worth noting the differences from the result of the
works \cite{don,BBDH}. In these papers, not only the gravitational field,
but also the scalar fields with the masses $M$ and $m$ are quantized.
In our opinion, this procedure is not logically consistent for the
interactions between macroscopic bodies. Here we follow an
alternative approach of \cite{DM97}, where the macroscopic bodies
are regarded as classical sources and only gravitons are quantum
fields. For this reason, the corresponding effective-quantum-gravity-based  potential analogous
to (\ref{pot_final}) does not have the ``classical'' $1/r^2$ term.
On top of this, the coefficient of the  ``quantum'' $1/r^3$ term is
slightly different in  \cite{DM97,EffQG-TNLM} compared to
\cite{BBDH}. On the other hand, both details have only marginal
relevance for the subject of our work. The reason is that
(\ref{pot_final}) with the coefficient (\ref{betinv}) is the
correction in the semiclassical case. The objects of quantization are
conformal fields which do not interact directly with the masses $M$
and $m$. Because of this fact, the mixed $P$-type diagrams of
\cite{don} do not contribute, and this explains the aforementioned
difference.

In the next sections we consider a qualitatively different
calculation of the gravitational potential, based on the weak quantum
correction to the Schwarzschild metric. Starting from this point,
we restrict our attention to the conformal matter fields only.

\section{Quantum conformal matter and trace anomaly}
\label{sec3}

Let us start with a brief review of the anomaly and the anomaly-induced
action. In addition, we derive some bulky equations which
will be used in what follows.

The anomaly-induced action is an economical and efficient way of
working with leading logarithms [e.g., the vacuum part of
Eq.~(\ref{EA-Weyl})] at the one-loop level. From the formal perspective,
the trace anomaly $\langle T^\mu_{\,\,\mu}\rangle $ is related to
the breaking of local conformal symmetry by the quantum corrections
\cite{CapDuf-74,duff77}. The equation for the induced action
has the form
\begin{align}
-\,\frac{2}{\sqrt{-g}} \,g_{\mu\nu}
\,\fdv{\Gamma_{\tx{ind}}}{g_{\mu\nu}}
\,\,=\,\,
\langle T^\mu_{\,\,\mu}\rangle
\,\,=\,\, -\, \big(\om C^2 + bE_4 + c\cx R\big) \,.
\label{eq4indAE}
\end{align}
The coefficients $\om$, $b$ and $c$ are the corresponding beta
functions from (\ref{Gadiv}) and (\ref{k123}). It is worth noting
that there is an ambiguity in the coefficient $c$ \cite{birdav,duff94},
which is related to the freedom to add a finite $R^2$-term to the
classical Lagrangian of the vacuum \cite{anomaly-2004}.
Eq.~(\ref{eq4indAE}) can be integrated  \cite{rie,frts84}.\footnote{
Recently, a slightly different way to arrive at the same result was
proposed in \cite{Kawai2025}.}
The anomaly-induced vacuum effective action is non-local,
but it can be localized by introducing two auxiliary scalar fields
$\vp,\psi$ \cite{a} (an alternative form was obtained in
\cite{Mazur:2001aa}),
\beq
\Gamma_{\tx{ind}} &=&
\cl{S}_c(g)
\,+\,
\int \dd[4]{x} \sqrt{-g}\,\Big\{
\,\frac12 \,\big( \vp \Delta_4 \vp - \psi\Delta_4\psi \big)
+ \big(l_1 \psi + k_1\vp\big) C^2
\nn
\\
&&
+ \,\, k_2 \ph \Big(E_4-\frac23 \square R\Big)
\Big\}
\,\, - \,\, \frac{3c+2b}{36}\int d^4x \sqrt{-g} \,R^2\,.
\label{Gaind}
\eeq
The coefficients $l_1,k_1,k_2$ are combinations of trace-anomaly coefficients,
\begin{align}
\label{eq:parameters-AISET}
k_1 = -\frac{\omega}{2\sqrt{-b}} \,, \quad
k_2=\frac{\sqrt{-b}}{2} \,, \quad
l_1=-k_1
\end{align}
and $\Delta_4$ is the Paneitz operator \cite{FrTs-superconf,Paneitz},
\begin{align}
\label{Paneitz-op}
\De_4
\,=\, \square^2 - 2R^{\mu\nu}\n_\mu \n_\nu
+\frac23 R\square - \frac13(\n^\mu R)\n_\mu \,.
\end{align}
The non-locality is hidden in the equations of motion for the
auxiliary fields,
\begin{subequations}
\label{eq:EoM-auxiliary-fields}
\begin{align}
\frac{1}{\sqrt{-g}} \fdv{\Gamma_{\tx{ind}}}{\vp}
&= \Delta_4 \vp + k_1 C^2 + k_2\qty(E_4 -\frac23 \square R) = 0 \,,
\\
\frac{1}{\sqrt{-g}} \fdv{\Gamma_{\tx{ind}}}{\psi}
&= -\Delta_4 \psi + l_1 C^2 = 0 \,,
\end{align}
\end{subequations}
Those are fourth-order differential equations, thus $\vp$ and $\psi$
depend on boundary conditions, which is a non-local effect.

Finally, the first term $\cl{S}_c(g)$ in the expression (\ref{Gaind})
is a conformally invariant action that cannot be fixed by the trace
anomaly. Consequently, this part of the effective action is not related
to the UV divergences and hence has no connection to the
logarithmic form factors. Since our interest is to explore the effect 
of the leading logarithmic form factors via trace anomaly, this term 
can be safely omitted. The final observation is that the  coefficient
of the $\int R^2$-term can be modified by adding a finite local term 
to the classical vacuum effective action, as is done in the 
Starobinsky inflationary model \cite{star}.

From the anomaly-induced effective action, one derives the reduced
stress-tensor
\begin{align}
\label{Smndef}
S_{\mu\nu}
\,=\,
\frac{2}{\sqrt{-g}} \,\fdv{\Ga_{\tx{ind}}}{g^{\mu\nu}} \,.
\end{align}
Here we used the notation $S_{\mu\nu}$ instead of the usual
$T_{\mu\nu}$ because here and in what follows we omit the
conformal term $\cl{S}_c(g)$.

In what follows, we shall need a more detailed form of  $S_{\mu\nu}$
compared to what was presented before \cite{balsan}. It proves useful
to separate the expression into the sum
\begin{align}
\label{Smn}
S_{\mu\nu}
\,=\, 
 K_{\mu\nu}(\vp) + K_{\mu\nu}(\psi) + E_{\mu\nu}(\vp)
 + C_{\mu\nu}(\vp) + C_{\mu\nu}(\psi) + L_{\mu\nu}\,.
\end{align}
In this sum, $K_{\mu\nu}$ is the tensor obtained by varying the part
of $\Gamma_{\tx{ind}}$ that is bilinear in the fields. $E_{\mu\nu}$
is the variation of the term containing the Gauss-Bonnet integrand, and
$C_{\mu\nu}$ arises from terms involving the square of the Weyl tensor. Finally, $L_{\mu\nu}$ is the contribution of the local
$\int R^2$ - term. This term does not depend on the auxiliary fields.
The symbolic calculations were performed using the
Mathematica xAct package \cite{Wolfram,xAct}. The results are as
follows,
\beq
K_{\mu\nu}(\vp)
&=&
\frac{2}{\sqrt{-g}}\fdv{}{g^{\mu\nu}}
\,\frac12 \int \dd[4]{x}\sqrt{-g} \vp \Delta_4 \vp
\nn
\\
&=&
\tfrac{1}{3} \big(g_{\mu \nu } R -2R_{\mu \nu} \big) (\na \vp)^2
- \tfrac{2}{3} (\na_{\al }\na_{\nu }\na_{\mu }\vp) (\na^{\al }\vp)
+ \tfrac{4}{3} (\na_{\al }\na_{\nu }\vp) (\na^{\al }\na_{\mu }\vp)
\nn
\\
&+&
\tfrac{3}{2} g_{\mu \nu } (\cx \vp)^ 2
+  \tfrac{5}{3} g_{\mu \nu } (\na^{\al }\vp) (\cx\na_{\al }\vp)
- 3 g_{\mu \nu } R_{\al \be } (\na^{\al }\vp) (\na^{\be }\vp)
\nn
\\
&+&
\tfrac{4}{3} R_{\mu \al \nu \be } (\na^{\al }\vp) (\na^{\be }\vp)
+\tfrac{1}{2} \vp R_{\mu \nu \al \be } (\na^{\be }\na^{\al }\vp)
- \tfrac{1}{3} g_{\mu \nu } (\na_{\al }\na_{\be }\vp)
(\na^{\al }\na^{\be } \vp)
\nn
\\
&-&
(\cx \na_{\nu }\vp) \na_{\mu }\vp
+ 3 R_{\nu \al } (\na^{\al }\vp) (\na_{\mu }\vp)
+ 3 R_{\mu \al } (\na^{\al }\vp) (\na_{\nu }\vp)
\nn
\\
&-&
\tfrac{2}{3} R (\na_{\mu }\vp) (\na_{\nu }\vp)
- 2 (\cx\vp) (\na_{\nu }\na_{\mu }\vp)
- (\cx \na_{\mu }\vp) (\na_{\nu }\vp)
\,.
\eeq
The same form holds for the action of $\psi$, with an
opposite sign, $K_{\mu\nu}(\psi)=-K_{\mu\nu}(\vp\to\psi)$.
Furthermore,
\beq
E_{\mu\nu}(\vp)
&=& \frac{2}{\sqrt{-g}}\fdv{}{g^{\mu\nu}}
\int \dd[4]{x} \sqrt{-g}\, k_2\Big(E_4-\frac23\square R\Big)\vp
\nn
\\
&=&
\,k_2 \Big[ -E_4  \vp g_{\mu \nu }
- 8 \vp R_{\mu }{}^{\al } R_{\nu \al }
+ 4 \vp R_{\mu \nu } R
- 8 \vp R^{\al \be } R_{\mu \al \nu \be }
\nn
\\
&+&
4 \vp R_{\mu }{}^{\al \be \ga } R_{\nu \al \be \ga }
- \tfrac{28}{3} R_{\mu \nu } \cx\vp
+ 4 g_{\mu \nu } R \cx\vp
+  \tfrac{4}{3} (\cx \na_{\nu }\na_{\mu }\vp)
\nn
\\
&+&
\tfrac{4}{3} (\na_{\al }R_{\mu \nu }) (\na^{\al }\vp)
- \tfrac{2}{3} g_{\mu \nu } (\na_{\al }R) (\na^{\al }\vp)
+  \tfrac{20}{3} R_{\nu \al } (\na^{\al }\na_{\mu }\vp)
\nn
\\
&+&
\tfrac{20}{3} R_{\mu \al } (\na^{\al }\na_{\nu }\vp)
- \,\tfrac{4}{3} g_{\mu \nu} \cx^ 2\vp
- 8 g_{\mu \nu } R_{\al \be } (\na^{\be }\na^{\al }\vp)
\nn
\\
&+&
\frac{32}{3} R_{\mu \al \nu \be } (\na^{\be }\na^{\al }\vp)
- \tfrac{4}{3} (\na^{\al }\vp) \na_{\mu }R_{\nu \al}
- \tfrac{4}{3}(\na^{\al }\vp)\na_{\nu }R_{\mu \al}
\nn
\\
&+&
\tfrac{2}{3} (\na_{\mu }R) (\na_{\nu }\vp)
+  \tfrac{2}{3} (\na_{\nu }R)(\na_{\mu }\vp)
- 4 R (\na_{\nu }\na_{\mu }\vp) \Big] \,,
\eeq
and
\beq
\,C_{\mu\nu}(\vp)
&=&
\frac{2}{\sqrt{-g}}\,\fdv{}{g^{\mu\nu}} \int \dd[4]{x} \sqrt{-g} \,k_1C^2\vp
\nn
\\
&=& \,\, k_1\Big[
g_{\mu \nu } C^2\vp
- 8 \vp R_{\mu }{}^{\al } R_{\nu \al }
+  \tfrac{4}{3} \vp R_{\mu \nu } R
+ 4 \vp R_{\mu }{}^{\al \be \la} R_{\nu \al \be \la }
- 4 R_{\mu \nu } \cx\vp
\nn
\\
&+& \,  \tfrac{4}{3} g_{\mu \nu } R \cx\vp
+ 4 \vp \cx R_{\mu \nu }
+ 8 (\na_{\al }R_{\mu \nu }) (\na^{\al }\vp)
+ 4 R_{\mu \al } (\na^{\al }\na_{\nu }\vp)
\nn\\
&+& \,\, 4 R_{\nu \al } (\na^{\al }\na_{\mu }\vp)
- \tfrac{2}{3} g_{\mu \nu } \cx R
- \tfrac{4}{3} g_{\mu \nu } (\na_{\al }R) (\na^{\al }\vp)
- 4 g_{\mu \nu } R_{\al \be } (\na^{\be }\na^{\al }\vp)
\nn\\
&-&  \,\,4 (\na^{\al }\vp) (\na_{\mu }R_{\nu \al })
+ 8 R_{\mu \al \nu \be } (\na^{\be }\na^{\al }\vp)
- 4 (\na^{\al }\vp) (\na_{\nu }R_{\mu \al })
\nn\\
&+&  \,\,\tfrac{2}{3} (\na_{\mu }\vp) (\na_{\nu }R)
+  \tfrac{2}{3} (\na_{\mu }R) (\na_{\nu }\vp)
- \tfrac{4}{3} R (\na_{\nu }\na_{\mu }\vp)
- \tfrac{4}{3} \vp (\na_{\nu }\na_{\mu }R) \Big] \,.
\eeq

The rule for the term with $\psi$ is
$C_{\mu\nu}(\psi)=C_{\mu\nu}(\vp\to\psi, k_1\to l_1)$.
To conclude, the last term is
\beq
\,L_{\mu\nu}
&=& \frac{2}{\sqrt{-g}}\,\fdv{}{g^{\mu\nu}} \int \dd[4]{x} \sqrt{-g}
\, \qty(-\tfrac{3c+2b}{36}) R^2
\nn
\\
&=& \,\, -\tfrac{3c+2b}{36} \qty[4 g_{\mu\nu} (\cx R)
+ 4 R R_{\mu\nu} -g_{\mu\nu} R^2 - 4(\n_\nu \n_\mu R) ] \,.
\label{Lmn}
\eeq

The expression for the stress-tensor \eqref{Smn} can be adapted
to a given vacuum state. For this, the equations of motion for the
auxiliary fields $\vp$ and $\psi$ should be solved and the solutions
replaced in (\ref{Smn}). This procedure uses the non-local nature
of the induced action \eqref{Gaind}. The boundary conditions for
the solutions of $\vp$ and $\psi$ enable one to use the information
about which vacuum state the quantum matter fields are in.

Although the trace anomaly \eqref{eq4indAE} is defined in terms of
curvature scalars, without dependence on the vacuum state,  the
applications of the anomaly-induced effective action may crucially depend
on the boundary conditions for the auxiliary fields, and these
conditions are different for different vacuum states
\cite{balsan,Mottola:2006ew}.  It was shown in \cite{balsan} that
with appropriate boundary conditions for $\vp$ and $\psi$, the
stress-tensor (\ref{Smndef}) can reproduce the leading behavior
of the vacuum average of the stress-tensor
$\langle T_{\mu\nu} \rangle$ of quantum fields in the Boulware
$|B\rangle$  and Unruh $|U\rangle$ vacuum states. For the
Hartle-Hawking vacuum $|H\rangle$ in a
Schwarzschild background, the situation is more complicated
\cite{balsan}, likely because this vacuum state is
thermal, which should modify the form of the anomaly and the
induced action.\footnote{I.Sh. is grateful to A.A. Starobinsky for
this observation.}

Since our goal is to explore the effect of the
modifications in the classical solution owing to the quantum effects
of matter fields, we assume that these modifications are described
 by the expression (\ref{Smndef}) and, consequently, may be
 dependent on the choice of particular solutions for the auxiliary
 fields $\vp$ and $\psi$.

\section{Newton potential from trace anomaly}
\label{sec4}

In this section, we consider the Boulware state, corresponding to an
observer situated far from the center of the black hole.  The
corresponding modifications in the Schwarzschild solution can
be analysed, in particular, in the large-$r$ limit. The results can
be compared to the ``traditional'' form of quantum corrections to
the Newtonian potential, described in the previous Sec. \ref{sec2}.
Thus,
let us derive the Newton potential from a gravitational action
\beq
\label{eq:EH+anomaly-induced-action}
&&
\cl{S}_g = \cl{S}_{\tx{EH}} + \Gamma_{\tx{ind}} \,,
\label{Gamma}
\eeq
where $\cl{S}_{\tx{EH}}$ is the first term in (\ref{classical-action})
and $ \Gamma_{\tx{ind}}$ is the induced action (\ref{Gaind}).

The Schwarzschild metric will be denoted as $\ga_{\mu\nu}$ and
considered as the zero-order, background approximation.
Following the effective approach, we will analyse the quantum
corrections perturbatively, treating $\Gamma_{\tx{ind}}$ as a
small addition to the Einstein-Hilbert term. Correspondingly,
the metric is a sum of the background and a small correction,
\beq
g_{\mu\nu}\,=\,\ga_{\mu\nu} \,+\,  \hbar h_{\mu\nu},
\label{hmn}
\eeq
where $\hbar$ is the parameter of the loop expansion. In what
follows, we restrict our discussion to linear order in this
parameter. On the other hand, we set $\hbar=1$ in all cases
when this does not create confusion. The equations of
motion follow from \eqref{eq:EH+anomaly-induced-action},
\begin{align}
\label{EeqsCor}
G_{\mu\nu} =\frac{\kappa^2}{2} S_{\mu\nu} \,,
\end{align}
where $G_{\mu\nu}$ is the Einstein tensor and $S_{\mu\nu}$ is
 given by \eqref{Smndef}. In the first order in $\hbar$,
these equations give (the arguments are indicated without indices)
\begin{align}
\label{eq:first-order-EoM}
G_{\mu\nu}^{(1)}(\ga,h;\,x)
\,=\, \int \dd[4]{y} \,\,
\frac{\de G_{\mu\nu}(\ga;\,x)}{\de \ga_{\al\be}(y)}\,\,h_{\al\be}(y)
\,=\, \frac{\kappa^2}{2} S_{\mu\nu}(\ga;\,x)\,,
\end{align}
where $G_{\mu\nu}^{(1)}(\ga,\,h)$ is the first-order expansion of
the Einstein tensor. Direct calculations in a generic background
$g_{\mu\nu}+h_{\mu\nu}$ give the expression
\beq
&&
G_{\mu\nu}^{(1)}(g,h)
\,=\, \frac12
\Big[ \n_\lambda \n_\mu h^\lambda{}_\nu
+ \n_\lambda \n_\nu h^\lambda{}_\mu -\n_\mu \n_\nu h
-\square h_{\mu\nu}
\nn
\\
&&
\qquad \qquad \quad
-\,\, g_{\mu\nu}\qty( \n_\alpha \n_\beta h^{\alpha\beta}
-\square h -R_{\alpha\beta} h^{\alpha\beta}) - R h_{\mu\nu} \Big]\,,
\label{G1}
\eeq
where all covariant derivatives are constructed with the background
metric. In the case of the Schwarzschild background, it reduces to
\beq
&&
G_{\mu\nu}^{(1)}(\ga, h)
\,=\,
\frac12\Big[ \n_\lambda \n_\mu h^\lambda{}_\nu
+ \n_\lambda \n_\nu h^\lambda{}_\mu -\n_\mu \n_\nu h
- \square h_{\mu\nu}
\nn
\\
&&
\qquad \qquad \quad
-\,\,
\ga_{\mu\nu}\qty( \n_\al \n_\be h^{\al\be} -\cx h) \Big] \,,
\label{Ggamma}
\eeq
where $\cx = \ga^{\mu\nu}\na_\mu \na_\nu$ and all indices
are raised and lowered with the background metric.

The quantum contribution to the stress-tensor $S_{\mu\nu}(\ga)$
is  proportional to $k_1$, $k_2$, $l_1$, which are all of the first
order in $\hbar$. Thus,  $S_{\mu\nu}(\ga)$  can be evaluated on
the classical Schwarzschild background $\ga_{\mu\nu}$ since any
$h_{\mu\nu}$-dependence produces terms $\cl{O}(\hbar^2)$.
The Schwarzschild background is a Ricci-flat spacetime, hence we
drop all terms proportional to $R_{\mu\nu}$ in the expressions for
$K_{\mu\nu}$, $E_{\mu\nu}$ and $C_{\mu\nu}$. Furthermore, the
tensor $L_{\mu\nu}$ in Eq.\  \eqref{Lmn} is identically zero in the
Schwarzschild background.
Another simplification in the tensors $K_{\mu\nu}$, $E_{\mu\nu}$ and
$C_{\mu\nu}$ is that all terms proportional to the auxiliary fields
(but not on their derivatives) are proportional to the tensor structure
\begin{align}
R_{\al\be\rho\si}^2 g_{\mu\nu}
- 4 R_{\mu}{}^{\al\be\ga} R_{\nu\alpha\beta\gamma} \,.
\end{align}
This combination can be verified to vanish in the Schwarzschild
metric case.
Therefore, the anomaly-induced stress-tensor $S_{\mu\nu}(\ga)$,
evaluated in the Schwarzschild background, depends only on
the derivatives of the auxiliary fields.

Independently of the mentioned simplifications, using the expression
for the stress-tensor requires solving the equations of motion for
the auxiliary fields  \eqref{eq:EoM-auxiliary-fields} and replacing
the solutions back into $S_{\mu\nu}(\ga)$.
The equations of motion \eqref{eq:EoM-auxiliary-fields} are
non-homogeneous differential equations. The freedom in the
choice of the homogeneous solution reflects the freedom to choose different Green functions in the non-local formulation
of the anomaly-induced effective action.
The solutions were
obtained in \cite{balsan} and enable one to interpolate between
different vacuum states. At this point, we note that the relation to
the classification of the vacuum state is in sharp contrast with
the ``traditional'' approach described in Sec.~\ref{sec2}, since in
the latter case the choice of the vacuum state is irrelevant.

In the Schwarzschild background, the Paneitz operator
\eqref{Paneitz-op} boils down to $\cx^2$ , while the square
of the Riemann tensor, in spherical coordinates, becomes
\beq
R_{\mu\nu\alpha\beta}^2=\frac{48 M^2}{r^6}\,.
\eeq
Thus, the equations for the auxiliary fields
\eqref{eq:EoM-auxiliary-fields} reduce to
\begin{align}
\label{eqauxSch}
\cx^2 \vp
\,=\, \frac{\alpha M^2}{r^6}
\,,
\quad
\mbox{where}
\quad
\alpha = -\,48(k_1+k_2) \,.
\end{align}
The equation for $\psi$ has coefficient $\al$ traded for
$\be =48l_1$.
As the structure of the equations is the same for both fields,
let us restrict the discussion to the $\vp$ case.

The general solution of \eqref{eqauxSch} is a particular solution
$\vp_p(r)$ plus the general homogeneous solution $\vp_h(r)$. The
homogeneous solution of the fourth-order differential equation
(\ref{eqauxSch}) is the sum of three different functions
$\vp_h^{(1)} + \vp_h^{(2)} + \vp_h^{(3)}$ summed with an
arbitrary constant. Since the stress-tensor in the Schwarzschild metric
depends only on the derivatives of the auxiliary fields, the derivative
$\vp'= \dv{\vp}{r}$ of \eqref{eqauxSch} is sufficient for
the calculation of $S_{\mu\nu}(s)$.

In addition to the radial-dependent solution, the auxiliary field
can depend on time $t$. Following the discussion in \cite{balsan},
we assume a linear time dependence,
\begin{align}
\vp(t,r) \,=\, \vp_t(t) \,\,+\,\, \mbox{function of $r$}\,,
\quad
\mbox{with}\,,
\quad
\vp_t(t) \,=\,  \frac{d}{2M}\,t
\end{align}
and $d$ is some dimensionless constant. The choice of a linear
time-dependence produces a static stress-tensor $S_{\mu\nu}$
with non-vanishing fluxes $S_{rt}\neq 0$  and possible
time-dependent $h_{\mu\nu}$.

All in all, the general solution for $\vp$ is
\beq
\vp(t,r) \,=\,  \vp_t(t) +\vp_h(r) + \vp_p(r) \,.
\label{timedep}
\eeq

One can use the solution of \cite{balsan}. Keeping the same
notations, we get
\begin{align}
\label{{solauxSch}}
\vp' = &
- \frac{A}{6} +\frac{B(r+r_g)}{3} -\frac{C}{r_g r}
- \frac{\alpha}{72M} +\frac{1}{r-r_g} \Big(\frac{Br_g^2}{3}
+ \frac{C}{r_g} -AM - \frac{\alpha}{24}\Big)
\notag
\\
&
+ \,\, \frac{r^3-r_g^3}{3r(r-r_g)}\,
\bigg(\frac{A}{r_g} -  \frac{\alpha}{12r_g^2}\bigg)
\log \Big(\frac{r}{r_g}-1\Big)
\notag
\\
&
 - \,\,\bigg[\frac{\alpha M}{18r(r-r_g)}
 + \qty(Ar_g-\frac{\alpha}{12})\frac{r^2}{3r_g^2(r-r_g)}\bigg]
 \log \Big(\frac{r}{r_g}\Big) \,.
\end{align}
$(A,B,C,d)$ are integration constants, and $r_g=2M$ is the
gravitational radius. The same solution holds for $\psi$ with new
integration constants, $(A,B,C,d) \to (A',B',C',d')$. These constants
have physical relevance since their values define the choice of the
vacuum states in which conformal fields, responsible for the trace
anomaly, are quantized \cite{balsan}.

The Newton potential corresponds to a static point-like
source generating a metric that is asymptotically Minkowski in the limit
$r \to \infty$. Obviously, the appropriate choice is the Boulware
vacuum state. This situation is approximated by setting all
constants to zero,
\beq
\label{ABCdzero}
(A,B,C,d) \,=\, (A',B',C',d') \,=\, 0 .
\eeq
The resulting stress-tensor $S_{\mu\nu}$ is diagonal and $h_{\mu\nu}$
is static. In the long-range limit $r\gg 2GM$, the stress-tensor with
(\ref{ABCdzero})  has the form
\beq
\label{eq:SET-Boulware-infinity}
S_\mu^{\,\,\nu} (\ga)
\,=\, \frac{1}{r^6}\,
\mqty( b_0 &0 &0 &0 \\ 0 &b_1 &0 &0
 \\ 0 &0 &b_2 &0 \\ 0 &0 &0 &b_3 ) \,,
\eeq
where $b_i$ are the following coefficients:
\beq
&&
b_0 = \, \frac{64}{81}
\big(185 k_1 k_2 + 133 k_2^2 \big) M^2 ,
\nn
\\
&&
b_1 = -\,\frac{b_2}{2}  = -\,\frac{b_3}{2}
= - \,\frac{128}{243}
\big( 29 k_1 k_2 + 55 k_2^2 \big) M^2 .
\label{b123}
\eeq
Eq.\ \eqref{eq:SET-Boulware-infinity} should be used as the source in
the \emph{r.h.s.} of Eq.~(\ref{eq:first-order-EoM}). After this, we
have to expand the \emph{l.h.s.}  for $r\gg 2M$ and arrive at the
correction to the classical metric and, eventually, to the Newtonian
potential.

Since the Einstein tensor is a second-derivative expression and
the source \eqref{eq:SET-Boulware-infinity} behaves as $\sim r^{-6}$,
from (\ref{eq:first-order-EoM}) it follows that, in the long-range
limit, the leading order of $h_\mu{}^\nu$ has to be $\sim r^{-4}$.
Thus, using the isotropic Newtonian form, we adopt the following
Ansatz for the metric perturbation $h_{\mu\nu}$:\footnote{We use
the spherical coordinates.}
\beq
\label{Newtonian-form}
h_{tt} = \frac{2 z_1}{r^4} \,,
\quad
h_{rr} = - \frac{2 z_2}{r^4} \,,
\quad
h_{\theta\theta} = 2 h_{rr} r^2 \,,
\quad
h_{\phi\phi} = 2 h_{rr} r^2 \sin^2\theta \,,
\eeq
where $z_1$ and $z_2$ are dimensionful constants that will be fixed
by the equations of motion \eqref{eq:first-order-EoM} in the large
radii limit.
 To find these constants, we note that the \emph{l.h.s.} of
the equations of motion is the expression \eqref{Ggamma}
supplemented with the Ansatz  (\ref{Newtonian-form}) and expanded
for $r\gg 2M$.
The \emph{r.h.s} is the source \eqref{eq:SET-Boulware-infinity}.
Solving for $z_1$ and $z_2$, one finds
\begin{align}
z_1 = \frac{(b_0+3b_1)\kappa^2}{48}\,,
\qquad
z_2 =-\frac{b_0 \kappa^2}{48}\,.
\end{align}

Now we are in a position to write the gravitational potential.
Following the notation of \eqref{Nmetric}, the function
$\Phi(r)$ for the theory \eqref{eq:EH+anomaly-induced-action}
is obtained from the relation $1+2\Phi=\ga_{00}+h_{00}$ as
a component of the metric with the one-loop correction.
Explicitly, the potential reads
\begin{align}
\label{New_iz_anom}
V(r) =  -\frac{GM}{r} +\frac{(b_0+3b_1)\pi G^3}{3 r^4} \,,
\end{align}
where we restored the dependence on the Newton's constant $G$ and we substituted $\kappa^2 = 16\pi G$.
The second term in this expression is induced by
$\Gamma_{\tx{ind}}$ and is proportional to $\hbar$, in addition we stress that the coefficient $(b_0+3b_1)$ is proportional to $M^2$.
It is worth noting that this result,
written in the Newtonian gauge for the metric, is independent,
in particular, of the gauge-fixing for the gauge vector field.

The result (\ref{New_iz_anom}) differs qualitatively from the
corrected Newtonian potential derived by another technique
(\ref{pot_final}), where the leading contribution was proportional
to $1/r^3$. Such a result was derived by the same method which was
developed in effective quantum gravity in
Refs.~\cite{DM97,EffQG-TNLM}.  The same $\mathcal{O}(1/r^3)$
form, albeit with a slightly different coefficient, was also found from
explicit computation of all loops including the ones with internal
lines of a massive scalar field (representing the macroscopic matter
source) \cite{BBDH,KirKhri}. From this perspective, using the
anomaly-induced effective action for deriving the quantum corrections
to the Newton potential leads to a surprising result since the
correction found in (\ref{New_iz_anom}) carries an
$\mathcal{O}(1/r^4)$ dependence, which is sub-leading
with respect to \eqref{pot_final}.

Trying to understand the origin of the difference, we note that the
$\mathcal{O}(1/r^4)$ correction comes from the dependence on the
choice of the vacuum state, something that does not exist in the
effective derivation in Sec.~\ref{sec2}. Of course, in effective
quantum gravity \cite{DM97,EffQG-TNLM} one is solving the
equations of motion in momentum space, assuming a flat background
in the weak-field limit. However, the choice of the vacuum state is
somehow more restrictive, in particular, the specific form of the
$r$-dependence looks more restrictive than the simple requirement
to have a Minkowski vacuum at spatial infinity.
The form of the correction in the potential that we obtain is a
consequence of the following two details:
\ \textit{i)} The
stress-tensor $S_{\mu\nu}$ that vanishes as $r^{-6}$ at spatial
infinity for the Boulware vacuum;
\ \textit{ii)} General relativity being a second-derivative theory of the
metric. These two features force the correction (\ref{New_iz_anom})
to behave as $r^{-4}$ at infinity, instead of $r^{-3}$.

The discrepancy related to the quantum corrections to the
Schwarzschild solution in the $r \gg 2M$ regime is not a completely
new issue. For example, motivated by the study of the $4D$ black
holes localized on the brane,  Anderson, Balbinot and Fabbri
\cite{ABF} derived the vacuum expectation value of the stress-tensor
induced by a quantum conformal scalar field in the Boulware state. The
calculation was based on direct renormalization in a specific
black hole background approach \cite{Anderson93,Anderson95}.
The result found in \cite{ABF} is that the stress-tensor behaves as
\beq
\expval{T_\mu{}^\nu} \sim \frac{1}{r^5}
\eeq
at spatial infinity. Assuming this asymptotic behaviour in the
general framework implies that the quantum-corrected Newtonian
potential behaves as  $r^{-3}$, i.e., it guarantees a qualitative
correspondence with the reliable result (\ref{pot_final}) from the
effective approach.

A natural question is whether the anomaly-induced stress-tensor
\eqref{Smn} is able to reproduce a leading $r^{-5}$ order in the
weak-field limit $r\gg 2M$ in the Schwarzschild spacetime with a
point-like mass $M$.
The tensor $S_{\mu\nu}(\ga)$ in the
Schwarzschild background depends on the set of auxiliary-field
constants $(A,B,C,d)$ and $(A',B',C',d')$, and on the trace anomaly
parameters $\om$, $b$, $c$. The latter appear in \eqref{Smn} through the
parameters  $k_1,k_2,l_1$ (see eqs.\ \eqref{eq:parameters-AISET}).
Is there a combination of these parameters that allows the
$r^{-5}$ asymptotic behavior of the stress-tensor?
We investigated the subject and found a negative answer, in a way
briefly described below.

We expanded the anomaly-induced stress-tensor
\eqref{Smn}, in the Schwarzschild background, in a power series in $r\gg 2M$,
\begin{align}
S_\mu{}^\nu(r) \sim a_0 + \frac{a_1}{r} + \frac{a_2}{r^2}
+ \frac{a_3}{r^3} + \frac{a_4}{r^4} +\frac{a_5}{r^5} \dots
\label{Smn-series}
\end{align}
Here $a_i=(a_i)_\mu{}^\nu$ are coefficient matrices, depending
on the parameters listed before. Let us mention that the unique
non-zero entries of the $a_i$ matrices are along the diagonal and the
$(a_i)_t{}^r, (a_i)_r{}^t$ components.
We report their expressions in Appendix \ref{app:appendix-coefficients} up to $a_6$. 
In what follows, however,
we will suppress the spacetime indices for simplicity of notation.

The stress-tensor \eqref{Smn-series} has the asymptotic behavior $r^{-5}$ if there is a combination of parameters $(A,B,C,d),(A',B',C',d')$ so that
\begin{align}\label{eq:parameters-ai=0}
a_0=a_1=a_2=a_3=a_4=0
\qquad
\mbox{and}
\qquad
a_5 \neq 0\,.
\end{align}
Eqs.\ \eqref{eq:parameters-ai=0} represent the system of equations
for the set of initial data $(A,B,C,d)$ and $(A',B',C',d')$,
depending on the given parameters $\om$, $b$, $c$, and $M$.
Looking at the full expressions in Appendix
\ref{app:appendix-coefficients}, one easily sees that
$a_0=a_1=a_2=0$ for
\begin{align}
\label{eq:system-boundary-constants-a0-a3-bis}
A^2=A'^2 \,, \qquad
B^2=B'^2 \,, \qquad
AB=A'B' \,, \qquad
Ad=A'd' \,.
\end{align}
The system \eqref{eq:system-boundary-constants-a0-a3-bis}
has solutions. Let us note that the coefficients $a_0,a_1,a_2$
are independent of parameters $k_1,k_2,l_1$, showing that the
corresponding terms in the series expansion are sourced by the
Paneitz-kinetic terms $\vp\Delta_4\vp$ and $\psi\Delta_4 \psi$
of \eqref{Gaind}.

If we add the constraints $a_3=a_4=0$ to the system, the
unique possible solution is
\begin{align}\label{eq:sol-a1=a2=a3=a4=0}
(A,B,C)=(A',B',C')=0 \,, \qquad d^2=d'^2  \,.
\end{align}
However, the solutions \eqref{eq:sol-a1=a2=a3=a4=0} also set to zero the diagonal components of $a_5$, while $(a_5)_t{}^r$ and $(a_5)_r{}^t$ are zero if one chooses $d=0$.
The choice $d=d'=0$ is physically motivated in order to have a static stress-tensor, as 
explained
in previous sections and first in \cite{balsan}.
Thus, we further assume
\begin{align}\label{eq:d=0}
d=d'=0 \,.
\end{align}

Altogether, based on the series expansion \eqref{Smn-series} with
coefficients listed in the Appendix, we have verified that it is not
possible to have an anomaly-induced stress-tensor behaving as
$r^{-5}$ at spatial infinity. Requirements for having a stress-tensor
$S_\mu{}^\nu \sim r^{-5}$ automatically set to zero the corresponding
coefficient $a_5$ promoting the stress-tensor to the subsequent
order. Therefore, the first non-zero order in the series expansion
\eqref{Smn-series} with these conditions is $a_6$, i.e.\ yielding
a stress-tensor
\begin{align}
S_\mu{}^\nu(\gamma) = \frac{a_6}{r^6} + \cl{O}\qty(r^{-7}) \,.
\end{align}
One can verify, using the formulas collected in Appendix, that there
are contributions to $a_6$ that are independent of the auxiliary-field
parameters while depending only on the trace-anomaly coefficients
$k_1$, $k_2$, and $l_1$.

The conditions \eqref{eq:sol-a1=a2=a3=a4=0}, \eqref{eq:d=0} are
those used for describing the Boulware vacuum. It is also worth
mentioning that the coefficients $a_3$, $a_4$, and $a_5$ written
in Appendix have a logarithmic dependence $\log(r/2M)$ that we
chose not to further approximate. They are not a problem for our
claim of having $S_\mu{}^\nu\sim r^{-6}$ since the terms having
the logarithmic dependence are set to zero by imposing  \eqref{eq:system-boundary-constants-a0-a3-bis}.

We now provide a qualitative explanation on why $a_6\neq 0$ when
conditions \eqref{eq:sol-a1=a2=a3=a4=0} are used. The stress-tensor
\eqref{Smn} depends on the fourth derivatives of the auxiliary
fields.
The equations of motion for the auxiliary fields, \eqref{eqauxSch},
are also of fourth order. Looking at \eqref{eqauxSch}, one sees that
the fourth derivative of $\vp$ is proportional to $\alpha M^2r^{-6}$,
independent of the boundary conditions $(A,B,C,d)$. And the same
holds for $\psi$, with $\beta M^2r^{-6}$. Such terms enter the
stress-tensor and survive the choice  \eqref{ABCdzero}. They
contribute to the $a_6/r^6$ term that we see in the weak-field
expansion $r\gg 2M$.
However, there is no such ``protection'' for the $1/r^5$-type
terms and hence they may vanish.

\section{Conclusions}
\label{sec5}

We performed the calculation of one-loop semiclassical corrections
to the Newton potential using two different approaches. In the first
case, one of the methods developed in the framework of effective
quantum gravity \cite{DM97,EffQG-TNLM} has been applied to
the loop contributions of the massless matter fields. The result fits
our intuitive expectations, in the sense that it has the same form as the
contributions of gravitons. The change concerns only the coefficients
of the leading term, while the functional dependence remains the
same $\mathcal{O}(1/r^3)$.

On the other hand, the second calculation produced an unexpected
result. We found the  semiclassical corrections to the Newton
potential as a long-distance limit of the $00$-component of the
quantum-corrected Schwarzschild solution. This correction to the
classical solution has to be considered on the basis of choosing
the vacuum state, in the framework of the anomaly-induced effective
action. The choice of the vacuum state is already a difference with
respect to the simplest calculation in the style of  effective
quantum gravity \cite{DM97,EffQG-TNLM}. 

The vacuum which admits taking the
limit $r \to \infty$ is the Boulware state. This state, and its
asymptotic properties at space infinity,  were previously used
in \cite{balsan} as part of the classification of vacuum states on
the basis of the nonlocal sector of the anomaly-induced effective
action of the vacuum.  The main feature, in this approach, is that
 the mean value of the energy-momentum tensor in the Boulware
vacuum state behave as $\mathcal{O}(1/r^6)$.

The $\mathcal{O}(1/r^3)$-type correction to
the Newtonian potential implies that the mean value of the
energy-momentum tensor in the given vacuum state should
behaves as $\mathcal{O}(1/r^5)$. Our analysis has shown that
there is no visible possibility to provide the ``standard''
$\mathcal{O}(1/r^3)$ correction to the potential with the ``standard''
$\mathcal{O}(1/r^6)$ asymptotic behavior of the energy-momentum
tensor. Thus, our results support the output of the recent
(compared to \cite{balsan}, where the anomaly-induced
framework for the classification of vacuum states was introduced)
analysis of \cite{ABF} and the much more recent one of
\cite{BoassoMaz2025} which also indicated the $\mathcal{O}(1/r^6)$
asymptotic behavior in the Boulware vacuum. One of the possible
explanations is that the $\mathcal{O}(1/r^5)$-terms are hidden in
the conformal term $\mathcal{S}_c$, however there is a more
general argument.

The anomaly-induced action is certainly the simplest representation of the loop corrections coming from the massless
conformal fields. The question is to which extent this form of
effective action reflects the real loop contributions. The
correspondence between leading one-loop logarithm and the
anomaly-induced action cannot be regarded as completely taken for granted,
as we showed in this work.
The way from the logarithmic form factor to the anomaly-induced
action goes through the dependence on the conformal factor of the
metric \cite{OUP}.
This may be a safe procedure in the case of a conformally
trivial metric, such as the behaviour of the conformal factor of the metric
in cosmology. However, the Schwarzschild metric is of a
different type, especially in the exterior part of the horizon, and
the conformal factor of the metric does not play such an important
role. This peculiarity explains our results (and the earlier ones of
\cite{ABF} and \cite{BoassoMaz2025}) concerning the
difference between the effective approach based on the  leading
one-loop logarithms and the one based on the anomaly.

\section*{Acknowledgements}

\noindent

The authors are grateful to R. Balbinot for encouraging
correspondence.
R.F.\ is grateful to Coordena\c{c}\~{a}o de Aperfei\c{c}oamento de
Pessoal de N\'{\i}vel Superior -- CAPES (Brazil) for supporting his
Ph.D. project.
The work of I. Sh. is partially supported by Conselho Nacional de
Desenvolvimento Cient\'{\i}fico e Tecnol\'{o}gico - CNPq under the grant
305122/2023-1.

\vskip 1.2cm

\paragraph{Note Added.}
After this paper was submitted to arXiv, we became aware of the work 
by Boasso and Mazzitelli \cite{BoassoMaz2025}, in which coherent 
results were obtained. Their findings further support the idea that the 
missing term required for agreement between the anomaly-induced 
effective action and the effective approach to gravity may be 
concealed within the conformal term $\mathcal{S}_c$.

\vskip 1.2cm


\section*{Appendix. Anomaly-induced stress-tensor in power series}
\label{app:appendix-coefficients}
We report here the coefficients of the series expansion \eqref{Smn-series} in $r\gg 2M$ of the anomaly-induced stress tensor in the Schwarzschild background.

The zeroth order:
\begin{align}
& (a_0)_t{}^t = \tfrac{25}{18} (B^2 -  B'^2) \,.
\nn\\
& (a_0)_r{}^r = \tfrac{47}{54} (B^2 -  B'^2) \,.
\nn\\
& (a_0)_\theta{}^\theta = (a_0)_\phi{}^\phi = \tfrac{47}{54} (B^2 -  B'^2) \,.
\nn\\
& (a_0)_t{}^r = (a_0)_r{}^t = 0\,.
\end{align}
%

First order:
\begin{align}
& (a_1)_t{}^t  = - \tfrac{2}{9} \bigl(10 A B
- 10 A' B' + (B^2 -  B'^2) M\bigr) \,.\nn\\
& (a_1)_r{}^r = \tfrac{10}{27}
\bigl(-6 A B + 6 A' B' + (B^2 -  B'^2) M\bigr) \,.\nn\\
& (a_1)_\theta{}^\theta = (a_1)_\phi{}^\phi
= - \tfrac{2}{27} \bigl(15 A B - 15 A' B' + (B^2 -  B'^2) M\bigr)
\,.\nn\\
& (a_1)_t{}^r = (a_1)_r{}^t = 0\,.
\end{align}
%

Second order:
\begin{align}
& (a_2)_t{}^t
= \tfrac{1}{18} \bigl(9 A^2 - 9 A'^2 - 24 A B M + 24 A' B' M
+ 4 (- B^2 + B'^2) M^2\bigr) \,.\nn\\
& (a_2)_r{}^r
= \tfrac{1}{6} \bigl(9 A^2 - 9 A'^2 - 24 A B M + 24 A' B' M
+ 4 (B^2 -  B'^2) M^2\bigr) \,.\nn\\
& (a_2)_\theta{}^\theta = (a_2)_\phi{}^\phi
= - \tfrac{2}{9} (B^2 -  B'^2) M^2 \,.\nn\\
& (a_2)_t{}^r = - (a_2)_r{}^t = - A d + A' d'\,.
\end{align}
%

Third order:
\begin{align}
& (a_3)_t{}^t  = - \tfrac{1}{27} M
\Bigl(27 A^2 + 40 A B M -  A' \bigl(27 A' + 40 B' M\bigr)\Bigr) \,.
\nn\\
& (a_3)_r{}^r = \tfrac{1}{81} \Bigl( 72 B C - 72 B' C' - 216 A k_2
+ 405 A^2 M - 405 A'^2 M \nn\\
& \hspace{2cm} - 216 d^2 M + 216 d'^2 M + 496 B' k_1 M
+ 584 A' B' M^2 + 96 B^2 M^3 - 96 B'^2 M^3 \nn\\
& \hspace{2cm} - 8 B M (46 k_1 + 46 k_2 + 73 A M)
- 96 (A B -  A' B') M^2 \log(\frac{r}{2 M})\Bigr) \,.\nn\\
& (a_3)_\theta{}^\theta
= (a_3)_\phi{}^\phi = - \tfrac{2}{81}
\Bigl(18 B C - 18 B' C' - 54 A k_2 + 81 A^2 M - 81 A'^2 M \nn\\
& \hspace{2cm} - 54 d^2 M + 54 d'^2 M + 124 B' k_1 M
+ 68 A' B' M^2 + 24 B^2 M^3 - 24 B'^2 M^3 \nn\\
& \hspace{2cm} - 4 B M \bigl(23 k_1 + 23 k_2 + 17 A M\bigr) - 24 \bigl(A B -  A' B'\bigr) M^2 \log\bigl(\frac{r}{2 M}\bigr)\Bigr) \,.\nn\\
& (a_3)_t{}^r = 0 \,.\nn\\
& (a_3)_r{}^t = 4 \bigl(A d -  A' d'\bigr) M\,.
\end{align}
%

Fourth order:
\begin{align}
& (a_4)_t{}^t  = \tfrac{1}{27} \Bigl(-63 A' C' + 18 B C M - 18 B' C' M - 160 A' k_1 M - 46 A^2 M^2 + 46 A'^2 M^2
\nn\\
& \hspace{2cm} - 54 d^2 M^2 + 54 d'^2 M^2 - 68 B k_1 M^2 + 220 B' k_1 M^2 - 68 B k_2 M^2 + 86 A' B' M^3
\nn\\
& \hspace{2cm} + 24 B^2 M^4 - 24 B'^2 M^4 + A (63 C + 272 k_1 M + 308 k_2 M - 86 B M^3) \nn\\
& \hspace{2cm} - 12 M^2 \bigl(7 A^2 + 2 A B M -  A' (7 A' + 2 B' M)\bigr) \log(\frac{r}{2 M})\Bigr) \,.
\nn\\
& (a_4)_r{}^r = \tfrac{1}{81} \Bigl(135 A' C' + 126 B C M - 126 B' C' M - 768 A' k_1 M + 942 A^2 M^2 - 942 A'^2 M^2 \nn\\
& \hspace{2cm} - 378 d^2 M^2 + 378 d'^2 M^2 + 1156 B k_1 M^2 + 292 B' k_1 M^2 + 1156 B k_2 M^2
\nn\\
& \hspace{2cm}
+ 890 A' B' M^3 + 168 B^2 M^4 - 168 B'^2 M^4 + A (-135 C + 528 k_1 M
+ 420 k_2 M\nn\\
& \hspace{2cm}
- 890 B M^3) + 12 M^2 \bigl(15 A^2 - 14 A B M + A' (-15 A' + 14 B' M)\bigr) \log(\frac{r}{2 M})\Bigr) \,.
\nn\\
& (a_4)_\theta{}^\theta
= (a_4)_\phi{}^\phi = \tfrac{1}{81} \Bigl(-135 A' C' - 90 B C M + 90 B' C' M
+ 768 A' k_1 M - 528 A^2 M^2
\nn\\
& \hspace{2cm}
+ 528 A'^2 M^2 + 270 d^2 M^2 - 270 d'^2 M^2 - 44 B k_1 M^2 - 44 B' k_1 M^2
- 44 B k_2 M^2 \nn\\
& \hspace{2cm}
- 358 A' B' M^3 - 120 B^2 M^4 + 120 B'^2 M^4 + A (135 C - 528 k_1 M - 528 k_2 M
\nn
\\
& \hspace{2cm}
+ 358 B M^3) + 60 M^2 \bigl(-3 A^2 + 2 A B M
+ A' (3 A' - 2 B' M)\bigr) \log(\frac{r}{2 M})\Bigr) \,.
\nn\\
& (a_4)_t{}^r = 0 \,.\nn\\
& (a_4)_r{}^t = 12 (A d -  A' d') M^2\,.
\end{align}
%

Fifth order:
\begin{align}
& (a_5)_t{}^t  = \tfrac{4}{27} M \Bigl(-27 A' C' + 18 B C M - 18 B' C' M - 300 A' k_1 M + 48 A^2 M^2 - 48 A'^2 M^2 \nn\\
& \hspace{2cm} - 54 d^2 M^2 + 54 d'^2 M^2 - 92 B k_1 M^2 + 52 B' k_1 M^2 - 92 B k_2 M^2 + 50 A' B' M^3 \nn\\
& \hspace{2cm} + 24 B^2 M^4 - 24 B'^2 M^4 + A (27 C + 240 k_1 M + 258 k_2 M - 50 B M^3) \nn\\
& \hspace{2cm} - 12 M^2 \bigl(3 A^2 + 2 A B M -  A' (3 A' + 2 B' M)\bigr) \log(\frac{r}{2 M})\Bigr)\,. \nn\\
& (a_5)_r{}^r = \tfrac{4}{135} M \Bigl(135 A' C' + 90 B C M - 90 B' C' M - 660 A' k_1 M \nn\\
& \hspace{2cm} + 720 A^2 M^2 - 720 A'^2 M^2 - 270 d^2 M^2 + 270 d'^2 M^2 \nn\\
 & \hspace{2cm} + 836 B k_1 M^2 + 116 B' k_1 M^2 + 836 B k_2 M^2 + 618 A' B' M^3 \nn\\
& \hspace{2cm} + 120 B^2 M^4 - 120 B'^2 M^4 - 3 A (45 C + 40 k_1 M + 70 k_2 M + 206 B M^3) \nn\\
& \hspace{2cm} + 60 M^2 \bigl(3 A^2 - 2 A B M + A' (-3 A' + 2 B' M)\bigr) \log(\frac{r}{2 M})\Bigr) \,.
\nn
\end{align}
\begin{align}
& (a_5)_\theta{}^\theta = (a_4)_\phi{}^\phi = - \tfrac{4}{135} M
\Bigl(135 A' C' + 90 B C M - 90 B' C' M - 930 A' k_1 M
\nn\\
& \hspace{2cm} + 495 A^2 M^2 - 495 A'^2 M^2 - 270 d^2 M^2 + 270 d'^2 M^2
\nn\\
& \hspace{2cm}
+ 44 B k_1 M^2 + 44 B' k_1 M^2 + 44 B k_2 M^2 + 362 A' B' M^3
\nn\\
& \hspace{2cm}
+ 120 B^2 M^4 - 120 B'^2 M^4 + A (-135 C + 690 k_1 M + 690 k_2 M - 362 B M^3)
\nn\\
& \hspace{2cm} + 60 M^2 \bigl(3 A^2 - 2 A B M + A' (-3 A' + 2 B' M)\bigr)
\log(\frac{r}{2 M})\Bigr)\,.
\nn\\
& (a_5)_t{}^r = 32 d (k_1 + k_2) M^2 \,.
\nn\\
& (a_5)_r{}^t = -32 M^2 \bigl(A' d' M + d (k_1 + k_2 -  A M)\bigr)\,.
\end{align}
%

Sixth order:
\begin{align}
& (a_6)_t{}^t = - \tfrac{2}{405} \Bigl(405 C^2 - 405 C'^2 - 9000 C' k_1 M - 1080 A' C' M^2 \nn\\
& \hspace{2cm} + 17280 k_1^2 M^2 + 30880 k_1 k_2 M^2 + 21920 k_2^2 M^2 + 1620 B' C' M^3 \nn\\
& \hspace{2cm} - 5520 A k_1 M^3 + 7440 A' k_1 M^3 - 6600 A k_2 M^3 \nn\\
& \hspace{2cm} - 4760 A^2 M^4 + 4760 A'^2 M^4 + 4860 d^2 M^4 - 4860 d'^2 M^4 \nn\\
& \hspace{2cm} + 8616 B k_1 M^4 - 24 B' k_1 M^4 + 8616 B k_2 M^4 + 3612 A B M^5 - 3612 A' B' M^5 \nn\\
& \hspace{2cm} - 2160 B^2 M^6 + 2160 B'^2 M^6 + 180 C M (58 k_1 + 58 k_2 + 6 A M - 9 B M^2) \nn\\
& \hspace{2cm} - 120 M^2 \bigl(9 A C - 9 A' C' + 12 A^2 M^2 + 2 A M (58 k_1 + 58 k_2 - 9 B M^2) \nn\\
& \hspace{2cm} - 2 A' M (50 k_1 + 6 A' M - 9 B' M^2)\bigr) \log(\frac{r}{2 M}) \nn\\
& \hspace{2cm} + 720 (A^2 -  A'^2) M^4 \bigl(\log(\frac{r}{2 M})\bigr)^2\Bigr)\,.
\nn
\end{align}
\begin{align}
& (a_6)_r{}^r  = - \tfrac{2}{243} \biggl(81 C^2 - 81 C'^2 - 1800 C' k_1 M - 1440 A' C' M^2 + 3456 k_1^2 M^2 \nn\\
& \hspace{2cm} - 1600 k_1 k_2 M^2 - 3392 k_2^2 M^2 + 540 B' C' M^3 + 8192 A k_1 M^3 \nn\\
& \hspace{2cm} + 2144 A' k_1 M^3 + 8840 A k_2 M^3 - 4544 A^2 M^4 + 4544 A'^2 M^4 \nn\\
& \hspace{2cm} + 1620 d^2 M^4 - 1620 d'^2 M^4 - 5640 B k_1 M^4 - 456 B' k_1 M^4 - 5640 B k_2 M^4 \nn\\
& \hspace{2cm} + 3828 A B M^5 - 3828 A' B' M^5 - 720 B^2 M^6 + 720 B'^2 M^6 \nn\\
& \hspace{2cm} + 36 C M \bigl(58 k_1 + 58 k_2 + 5 M (8 A - 3 B M)\bigr) \nn\\
& \hspace{2cm} - 24 M^2 \Bigl(9 A C - 9 A' C' + 80 A^2 M^2 + 2 A M (58 k_1 + 58 k_2 - 15 B M^2) \nn\\
& \hspace{2cm} - 10 A' M \bigl(10 k_1 + M (8 A' - 3 B' M)\bigr)\Bigr) \log(\frac{r}{2 M}) \nn\\
& \hspace{2cm} + 144 (A^2 -  A'^2) M^4 \bigl(\log(\frac{r}{2 M})\bigr)^2\biggr)\,.
\nn
\end{align}
\begin{align}
& (a_6)_\theta{}^\theta = (a_6)_\phi{}^\phi =
\tfrac{4}{1215} \Biggl(405 C^2 - 405 C'^2 - 9000 C' k_1 M - 5040 A' C' M^2 \nn\\
& \hspace{2cm} + 17280 k_1^2 M^2 - 8000 k_1 k_2 M^2 - 16960 k_2^2 M^2 \nn\\
& \hspace{2cm} + 1890 B' C' M^3 + 8260 A k_1 M^3 + 10420 A' k_1 M^3 + 8260 A k_2 M^3 \nn\\
& \hspace{2cm} - 9250 A^2 M^4 + 9250 A'^2 M^4 + 5670 d^2 M^4 - 5670 d'^2 M^4 \nn\\
& \hspace{2cm} - 588 B k_1 M^4 - 588 B' k_1 M^4 - 588 B k_2 M^4 + 7494 A B M^5 - 7494 A' B' M^5 \nn\\
& \hspace{2cm} - 2520 B^2 M^6 + 2520 B'^2 M^6 + 90 C M \bigl(116 k_1 + 116 k_2 + 7 M (8 A - 3 B M)\bigr) \nn\\
& \hspace{2cm} - 120 M^2 \biggl(56 A^2 M^2 + A (9 C + 116 k_1 M + 116 k_2 M - 21 B M^3) \nn\\
& \hspace{2cm} -  A' \Bigl(9 C' + M \bigl(100 k_1 + 7 M (8 A' - 3 B' M)\bigr)\Bigr)\biggr) \log(\frac{r}{2 M}) \nn\\
& \hspace{2cm} + 720 (A^2 -  A'^2) M^4 \bigl(\log(\frac{r}{2 M})\bigr)^2\Biggr)\,.
\nn
\end{align}
\begin{align}
& (a_6)_t{}^r = 0 \,.
\nn
\\
& (a_6)_r{}^t = -16 M^3 \bigl(5 A' d' M + d (8 k_1 + 8 k_2 - 5 A M)\bigr) \,.
\end{align}
%


\end{document}